# Electrical Gating of the Charge-Density-Wave Phases in Quasi-2D $h$-BN/1T-TaS$_2$ Devices

Maedeh Taheri[1], Jonas Brown[1], Adil Rehman[2], Nicholas R. Sesing[3], Fariborz Kargar[1,4], Tina T. Salguero[3], Sergey Rumyantsev[2], and Alexander A. Balandin[1,4*]

[1]Nano-Device Laboratory, Department of Electrical and Computer Engineering, Bourns College of Engineering, University of California, Riverside, California 92521 U.S.A.

[2]CENTERA Laboratories, Institute of High-Pressure Physics, Polish Academy of Sciences, Warsaw 01-142 Poland

[3]Department of Chemistry, University of Georgia, Athens, Georgia 30602 U.S.A.

[4]Phonon Optimized Engineered Materials Center, Materials Science and Engineering Program, Bourns College of Engineering, University of California, Riverside, California 92521 U.S.A.

---

* Corresponding author (A.A.B.): balandin@ece.ucr.edu ; web-site: http://balandingroup.ucr.edu/

**Abstract**

We report on electrical gating of the charge-density-wave phases and current in *h*-BN capped three-terminal 1T-TaS$_2$ heterostructure devices. It is demonstrated that the application of a gate bias can shift the source-drain current-voltage hysteresis associated with the transition between the nearly commensurate and incommensurate charge-density wave phases. The evolution of the hysteresis and the presence of abrupt spikes in the current while sweeping the gate voltage suggest that the effect is electrical rather than self-heating. We attribute the gating to an electric-field effect on the commensurate charge-density-wave domains in the atomic planes near the gate dielectric. The transition between the nearly commensurate and incommensurate charge-density-wave phases can be induced by both the source-drain current and the electrostatic gate. Since the charge-density-wave phases are persistent in 1T-TaS$_2$ at room temperature, one can envision memory applications of such devices when scaled down to the dimensions of individual commensurate domains and few-atomic plane thicknesses.

The charge-density-wave (CDW) phase is a macroscopic quantum state consisting of a periodic modulation of the electronic charge density accompanied by a periodic distortion of the atomic lattice [1-3]. The early work on CDW effects, performed with bulk samples of the quasi-one-dimensional (1D) metallic crystals, revealed many spectacular phenomena — nonlinear electron transport, oscillating electric current for constant voltages, giant dielectric response, and multi-stable conducting states [1-4]. Recent years witnessed a rebirth of the field of CDW materials and devices, partially driven by an interest in layered quasi-2D van der Waals materials where CDW phases can manifest themselves at room temperature (RT) and above [5-21]. The size and geometry of quasi-2D CDW films provide new opportunities for device fabrication. The first reports of de-pinning of CDWs in quasi-2D materials have emerged suggesting common features among CDW phenomena in quasi-1D and quasi-2D systems [22-23]. However, there is also an understanding of differences in physics governing CDW phases in material systems of different dimensionalities and crystal structures [24-25].

The 1T polymorph of TaS$_2$ is one of the prominent members of the quasi-2D van der Waals materials of the transition-metal dichalcogenide group that reveals several CDW phases and transitions between them, in the form of resistivity changes and hysteresis [10, 14-15, 18-21]. Two of the phase transitions in 1T-TaS$_2$ are above RT. These transitions can be induced by temperature, electric current, light, and other stimuli. It was noted in the prior works that 1T-TaS$_2$ CDW devices have the potential for application in oscillators, detectors, and radiation-hard electronics [21, 26-28]. The devices reported to date, were *two-terminal*, with the resistive state and hysteresis controlled by the source-drain current *via* the voltage-controlled resistor connected in series [21, 27]. These CDW devices utilized the hysteresis window and negative feedback for their operation rather than the on-off ratio in the resistance change. There have been reports that described CDW transitions in 1T-TaS$_2$ induced by visible light or electromagnetic radiation in MHz – GHz frequency range [29-31]. The question of the role of local heating in CDW devices, as a result of passing current, remains open. It likely depends on the specific device design and the speed of switching [18, 21, 32-34]. Despite numerous attempts, the electrical gating of the CDW phase remains elusive. Apart from the demonstrated possibility of modulating 1T-TaS$_2$ properties *via* ion intercalation in chemical field-effect transistors [15], we are not aware of any successful demonstration of changing the source-drain current and the current hysteresis by the application of a gate bias. Gating of CDW phases, particularly near RT, would substantially enhance the functionality of the CDW devices and open novel application domains, *e.g.* in non-volatile memory. Electrical gating is also important for answering a fundamental science question: "Can one achieve an electrical switching of CDW phase *via* the pure field effect, without any local Joule heating involved?"

The main difficulty of electrical gating of the CDW phases and currents in 2D van der Waals materials is associated with the fact that different phases still have a rather high concentration of charge carriers. Below 550 K, 1T-TaS$_2$ is in the metallic-like incommensurate CDW (IC-CDW) phase. The nearly commensurate phase (NC-CDW) appears below 350 K and persists approximately until 180 K. Below this temperature, 1T-TaS$_2$ enters the commensurate CDW (C-CDW) phase. In the NC-CDW phase, the hexagonal-shaped C-CDW clusters are separated by the incommensurate regions. As the temperature increases to 350 K, the domains melt, and the material falls into the IC-CDW state. The transition from the NC-CDW phase to the IC-CDW phase reveals itself as a change in the resistivity of the device channel accompanied by hysteresis. The carrier concentrations in these two CDW states are high: $10^{21}$ cm$^{-3}$ and $10^{22}$ cm$^{-3}$ for the higher resistive NC-CDW and lower resistive IC-CDW phases, respectively [15, 35]. The high concentration of charge carriers, *N*, results in a small relative change of the carrier concentration, *δN/N*, due to the large *N* and strong screening of the gate potential. Reducing the thickness of 1T-TaS$_2$ thin film for a stronger gating effect is not necessarily a viable approach because at small thicknesses the CDW phases can be locked and some phase transitions disappear [15, 36-37]. It is known that the NC-CDW – C-CDW transition disappears when the 1T-TaS$_2$ channel thickness decreases below ~9 nm [15, 37]. On the other side, the incommensurate regions, also referred to as the discommensurate network, may be rather narrow in the NC-CDW phase. The latter may reduce the screening of the electrical field created by the gate, allowing for its deeper penetration to the channel layer. The above considerations can be relevant to other quasi-2D CDW van der Waals materials.

In this Communication, we report the first successful demonstration of the electrostatic gating of the current in the NC-CDW phase near RT, and the electric field switching of the CDW phases in quasi-2D CDW van der Waals materials. It is demonstrated that the hysteresis in the source-drain current-voltage (I-V) characteristics that accompany the transition between the NC-CDW and IC-CDW phases can be shifted by applying a gate bias. The effect is electrical rather than self-heating. We also show that the transition between the NC-CDW and IC-CDW can be induced both by the source-drain current and the electrostatic gate.

## Results

**Device fabrication and testing.** The bulk 1T-TaS₂ crystals for this study were grown by the chemical vapor transport (CVT) method [16, 18, 38-41]; structure and composition were confirmed by X-ray diffraction and energy dispersive spectroscopy, respectively. The thin films of the material were mechanically exfoliated from the small 1T-TaS₂ crystals and transferred onto Si/SiO₂ substrates. For fabricating the device channels, we selected thin films with thicknesses ranging from ~15 nm to ~35 nm. The layers of 1T-TaS₂ were capped with electrically insulating thin films of *h*-BN with a few nm thicknesses for protection from environmental exposure and degradation [21, 37, 42]. This design also improves heat dissipation from the 1T-TaS₂ channel owing to the high thermal conductivity of *h*-BN. For comparison, the bulk thermal conductivity of 1T-TaS₂ in the NC-CDW state is ~ 4 W/mK to 6 W/mK while that of a few-layer *h*-BN is ~ 360 W/mK [18, 43-44]. We have also experimented with a more sophisticated design where two layers of *h*-BN were used so that the 1T-TaS₂ channel is completely encapsulated with *h*-BN. Figure 1a shows the schematic of h-BN-encapsulated 1T-TaS₂ devices. The performance of *h*-BN capped and *h*-BN encapsulated devices was similar and discussed jointly in the text below. The source-drain contacts were patterned with the electron beam lithography followed by the atomic layer etching step to remove the *h*-BN layer underneath the contacts. The evaporation of Ti / Au with 10 nm / 100 nm thicknesses was used to fabricate device structures with the lateral dimensions of 1 μm – 3 μm by 1 μm – 3.5 μm. The back gate consisted of a silver contact deposited on the highly doped Si substrate. The 300-nm thick layer of SiO₂ acted as the gate dielectric. An optical microscopy image of a representative *h*-BN/1T-TaS₂/*h*-BN device structure is presented in Figure 1b.

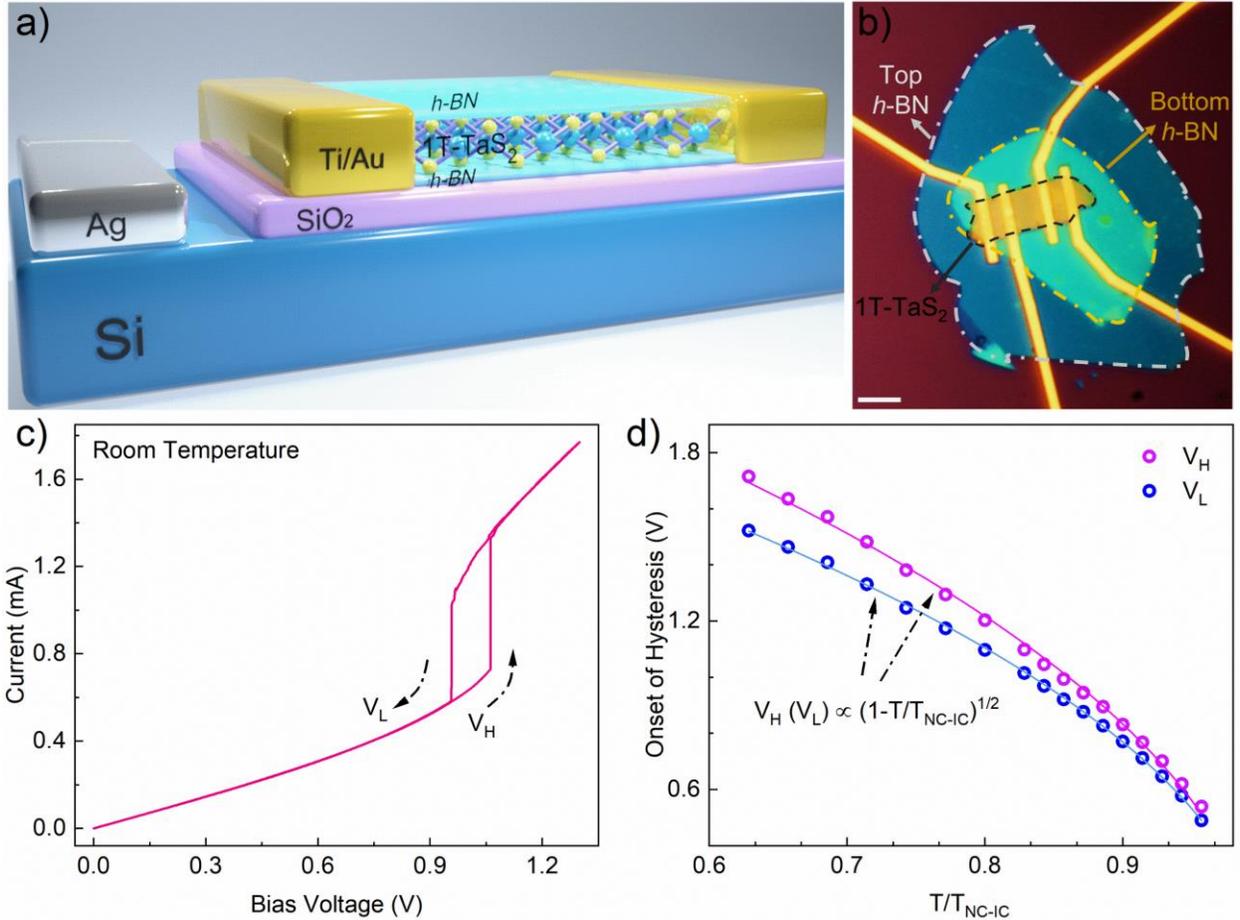

**Figure 1: Device structure and current-voltage characteristics.** (a) Schematic of 1T-TaS₂ heterostructure devices encapsulated in *h*-BN layers. For some devices, only the top h-BN capping layer was used. In 1T-TaS₂ crystal, the Ta atoms are depicted in blue while sulfurs are in yellow colors. (b) Optical image of the fabricated device with the channel thicknesses of ~35 nm. The scale bar is 5 μm. (c) I-V characteristics of 1T-TaS₂ device at room temperature. The arrows indicate the direction of source-drain bias sweep. (d) Dependency of the hysteresis onset voltages, $V_H$ and $V_L$, on temperature. The theory fitting $(1-T/T_{NC-IC})^{1/2}$ is shown with the purple and blue curves for $V_H$ and $V_L$, respectively. The measured $T_{NC-IC}$ for this device is 350 K.

We have tested about 20 different $h$-BN/1T-TaS$_2$ devices with varying channel thickness, $H < 35$ nm, and source-drain separation, $L$. Typical I-Vs with the signature hysteresis, measured at RT for a device with a channel thickness of ~ 15 nm, can be seen in Figure 1c. The voltage sweep direction is indicated by arrows. The hysteresis is due to the NC-CDW to IC-CDW phase transition. Measured as a function of temperature at small electric biases, *i.e.,* near-equilibrium conditions, this transition occurs at 350 K. In our measurements, the ambient temperature was kept constant, and the transition was triggered by applying the source-drain voltage and passing a current through the channel. Our prior studies for 1T-TaS$_2$ devices on Si/SiO$_2$ substrates, without $h$-BN capping, indicated that the local Joule heating produced by the current is essential for inducing the phase transition [18, 32]. The temperature dependence of the threshold voltages, $V_H$ and $V_L$, follows the relation $(1-T/T_{NC-IC})^{1/2}$ (see Figure 1d). This temperature dependence agrees with prior studies and is explained by the theories of the CDW order parameter [21]. The hysteresis part of the I-V provides functionality for CDW devices such as voltage-controlled oscillators that can operate at RT [21, 29]. A possibility of affecting the hysteresis with the electrostatic gate or changing the electrical current with the gate voltage in the linear part of the I-Vs would substantially increase the functionality of CDW devices.

**Electrical gating of CDW phases and current.** First, we attempted to change the I-Vs by applying the gate bias, $V_g$, in the range from -20 V to 40 V. The values of the gate bias are rather large due to the back-gate design and thick layers of SiO$_2$ dielectric. As mentioned above, it is unlikely that one can change the total current substantially, owing to the high carrier concentration and the large thickness of the samples. The collective CDW current in 2D materials is also small [45]. There is no well-defined de-pinning and sliding of CDW in 2D materials similar to that in metals with quasi-1D crystal structure, which would result in a strong increase in the total current and emergence of the AC component, termed "narrow-band noise" [1]. The latter is likely due to the domain structure of the NC-CDW phase in 2D materials. In the NC-CDW phase, the domains or islands of the C-CDW phase, which consist of atoms arranged in the "David-star" formation, are separated by the metallic IC-CDW phase, *i.e.* discommensurate regions. The C-CDW phase of 1T-TaS$_2$ has a band gap of about 100 meV to 200 meV [46-47]. How the size of the islands affects the bang-gap is not well understood. The IC-CDW phase surrounding the C-CDW islands has an electrical conductivity of at least an order of magnitude higher [37]. It is possible that the electric field applied perpendicular to the channel can produce some effect on the C-CDW domain structure in a few 1T-TaS$_2$ atomic planes, closest to the gate, thus altering the channel current. If the discommensurate regions have narrow width compared to the size of the C-CDW islands, it is possible that the screening is not effective, and the electric field of the gate penetrates deeper into the channel.

Figure 2a and 2b show the I-V characteristics near the hysteresis window for two different devices at the temperatures of 240 K and 200 K, respectively. These temperatures were selected somewhat above the C-CDW transition and below RT to ensure that the devices are in the NC-CDW phase. The data in Figure 2a are for the thinner channel ($H \sim 15$ nm) while the data in Figure 2b are for the thicker channel ($H \sim 35$ nm) device. One can see that we succeeded in shifting the hysteresis region by the applied gate bias. The changes in the hysteresis, characterized by the on-set voltages, $V_L$ and $V_H$, are non-monotonic with the gate bias, $V_g$. The changes are only partially reversible. This suggests that the electric field of the gate produces some changes to the C-CDW domains or their ordering. The semiconducting domains can be charged with the electrons from the surrounding IC-CDW phase or shifted or, possibly, merged with other domains when the field is strong enough. These mechanisms can produce permanent changes so that the hysteresis I-Vs do not return to the same positions when a reverse gate bias is applied. In measuring these I-Vs under different gate biases, we wanted to exclude or minimize the local heating effects. For fixed gate biases, we swept the source-drain voltage with an interval of $\sim 10$ min to avoid heat accumulation in the device structure. One can see in both Figures 2a and 2b that the area of the hysteresis remains unchanged at different gate biases. This is an indication that the shift of the hysteresis with the gate bias, *i.e.*, $V_H(V_g)$ and $V_L(V_g)$, is not due to the heating but rather due to the direct electric field effect. The hysteresis changes due to local heating are accompanied by the decreasing hysteresis size [21]. The hysteresis is large at low temperatures, and it collapses at T $\sim 350$ K, which is the NC-CDW – IC-CDW phase transition temperature near equilibrium, *i.e.*, measured at small biases.

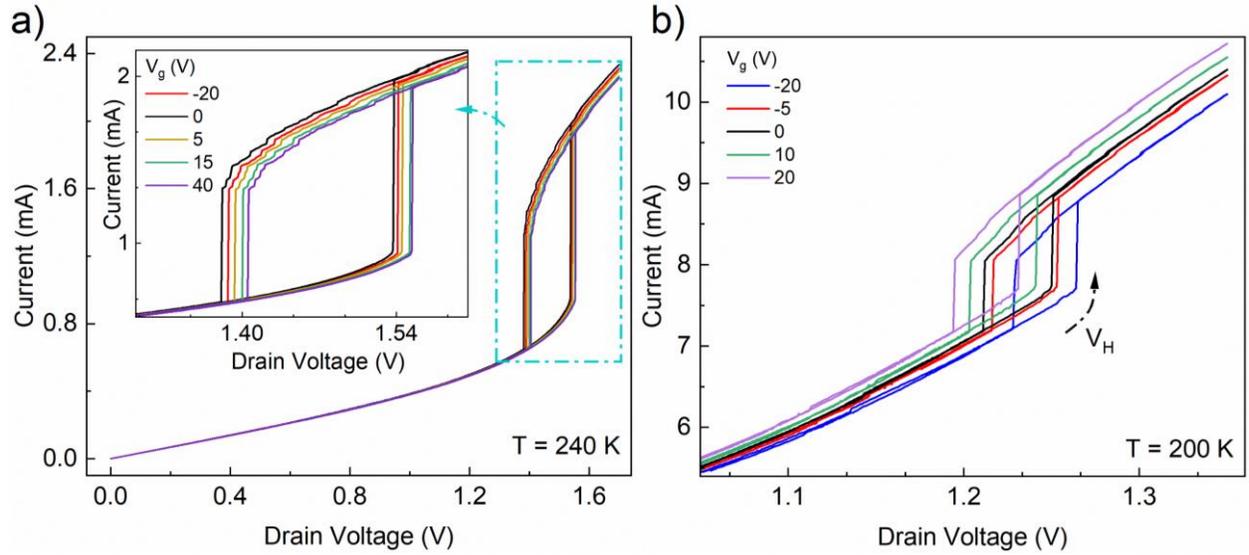

**Figure 2: Electrical gating of the charge-density-wave hysteresis.** (a) Drain-source I-V characteristics of the *h*-BN capped 1T-TaS$_2$ device ($H \sim 15$ nm) under the fixed gate biases $V_g$, at T = 240 K. Note that the linear region and the size of hysteresis remain unchanged under the gate bias. The inset shows the enlarged hysteresis, depicting a non-monotonic shift in the values of the threshold voltages, $V_H$ and $V_L$, which define the boundaries of the hysteresis, as the gate bias, $V_g$, varies from -20 V to 40 V. (b) Drain-source I-V characteristics of another device ($H \sim 35$ nm), demonstrating the shift in the threshold voltages $V_H$, as the $V_g$ takes values from -20 V to 20 V at T = 200 K. The drain bias sweeps have been performed with sufficient time intervals to avoid heat accumulation.

To better understand the mechanism of gating the CDW phases and currents, we measured the current while sweeping the gate voltage, $I_d(V_g)$, at T = 240 K. The temperature was intentionally selected in the middle of the NC-CDW phase. For these measurements, we fixed the source-drain voltage, $V_d$, and swept the gate voltage with the constant speed of 500 mV/s. We waited for a sufficient time between the voltage sweeps to avoid charge or heat accumulation. To analyze the data, we divided the I-V characteristics into three distinctive regions (see Figure 3a). Region I is the linear segment of I-Vs; region II is where the I-Vs become super-linear and evolve into the NC-CDW – IC-CDW phase transition; and region III is in the vicinity of $V_H$ and further away into the metallic IC-CDW phase. At low drain bias, the CDW domains are pinned, and electrical conduction is mostly carried through the IC-CDW metallic regions surrounding the CDW islands.

In this transport regime, the application of the gate bias does not produce much change (Figure 3b). At the lowest drain voltages $V_d = 0.03$ V and $V_d = 0.13$ V, one cannot observe any change in $I_d(V_g)$ at all. As $V_d$ increases, one can start noticing small spikes in the current at certain gate biases, which likely correspond to de-pinning or other changes in the C-CDW domains. The C-CDW domains can be viewed as semiconducting owing to the band gap opening [15, 37, 48].

Figure 3c shows current traces while sweeping the gate voltage in region II where the source-drain I-Vs become super-linear. This is the transport regime, where domains de-pin and can be changed by the electrical gate more efficiently. The variations in the current level become close to ~ 15 µA in the super-linear transport regime, *i.e.,* region II, as compared to ~ 5 µA in the linear regime, *i.e.,* region I. In Figure 3d, one can see a much stronger effect of the gate near the NC-CDW – IC-CDW phase transition region. The fluctuation in the source-drain current reaches the level of ~ 2 mA, which is on the order of the change in current when the system transitions between the NC-CDW and the IC-CDW phases. At $V_d = 1.64$ V the 1T-TaS₂ channel is entering the IC-CDW phase, but it is still unstable near the transition point, $V_H$. The sweeping of the gate bias produces the strongest effect here. At the gate bias in the vicinity of the NC-CDW – IC-CDW phase transition, there is local heating owing to the source-drain current. However, the abrupt variations in the current are due to the application of the gate voltage which changes the domain structure in the channel. As the drain bias increases further, $V_d \geq 1.75$ V, the channel becomes fully metallic in the IC-CDW phase, and the gate action stops. In this phase, the high concentration of charge carriers leads to strong screening and small $\delta I/I \sim \delta N/N$ variation.

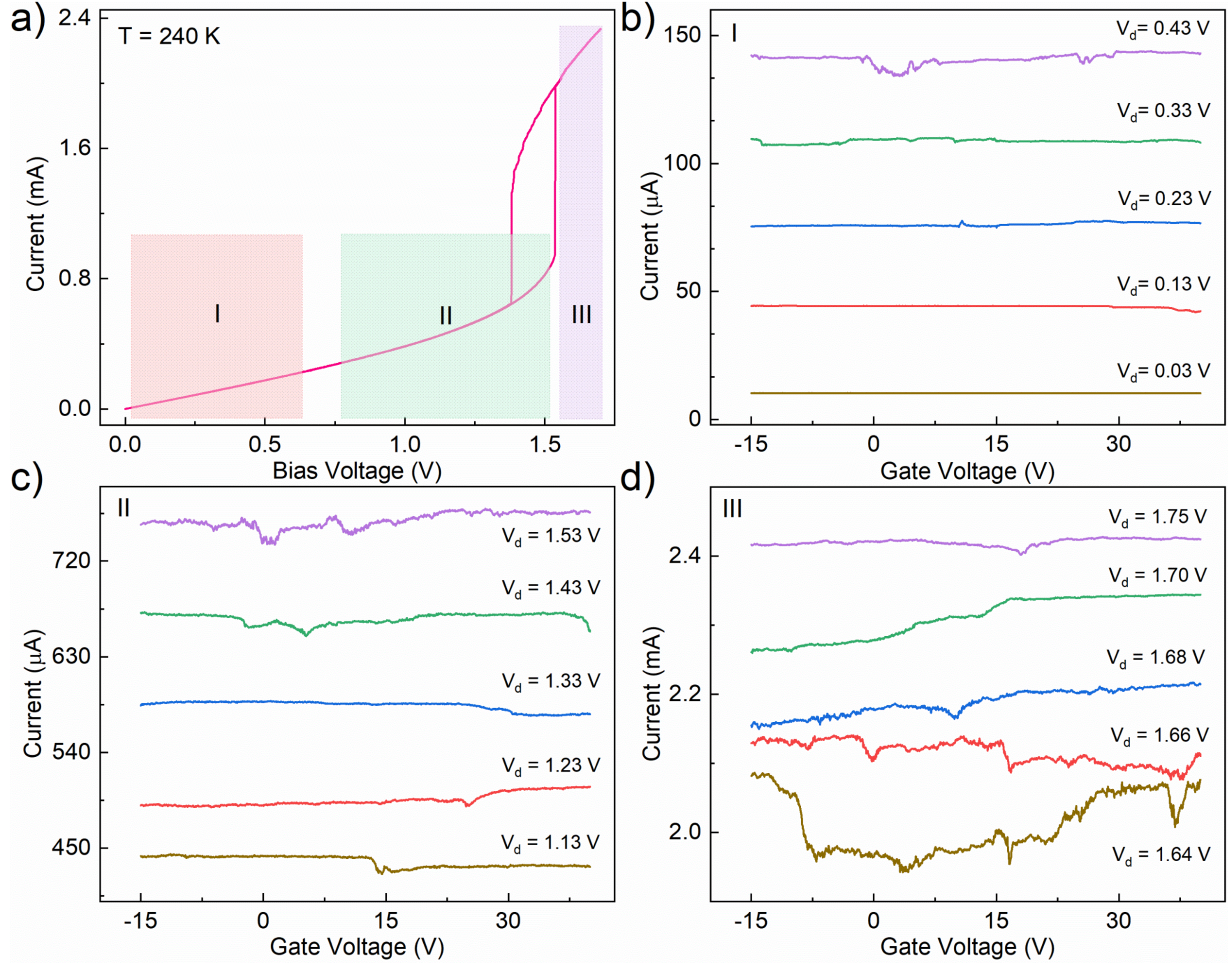

**Figure 3: Changing the charge-density-wave current with the gate.** (a) Measured I-V characteristics of the *h*-BN/1T-TaS₂ heterostructure device, without applied gate bias, at 240 K. The shaded regions I, II and III represent linear, super-linear and near-$V_H$ IC-CDW regions, respectively. The current traces measured with the gate sweep rate of 500 mV/S for fixed $V_d$ values for three regions are shown in (b), (c) and (d) panels, respectively. The current variations are negligible in region I as seen in (b). The current spikes grow in region II (c), and achieve maximum in region III (d).

While the trend described above is clear and consistent for the tested devices, one may notice a small offset between the hysteresis $V_L$ and $V_H$ values (see Figure 3a) and the changes in the transfer characteristics (see Figure 3d). For example, the strongest change in the transfer characteristics happens at $V_d = 1.64$ V while this value is already slightly larger than $V_H$, *i.e.* it is outside of the hysteresis in Figure 3a. This is because the I-V characteristics are taken at zero gate bias and in one sweep. However, the characteristics presented in Figure 3 (b-d) are taken at a fixed source-drain bias and an additional field produced by the gate. The latter explains why the hysteresis window, defined by $V_L$ and $V_H$, during the transfer characteristic measurements can be somewhat shifted from that presented in Figure 3a at $V_g = 0$ V. One can also argue that the C-CDW domains are not fully melted at $V_d$ slightly above $V_H$, and current variations can still be observed. These points are addressed in more detail in the discussion below.

We now consider in more detail the possibility of switching the system between the NC-CDW and IC-CDW phases, illustrated in Figure 4a, by application of a gate bias. Figure 4b shows the drain-source I-V characteristics of another device for two fixed gate voltages, $V_g$. The measurements were performed at the temperature of 210 K to ensure that the system is in the NC-CDW phase. As the $V_g$ switches from 20 V to 35 V, the hysteresis threshold voltage, $V_H$, shifts to a higher value. For a selected $V_d$ near 1.8 V ($V_d = 1.83$ V in Figure 4b), the channel is in the vicinity of the phase transition point and the suitable $V_g$ bias can switch the channel between the NC-CDW and IC-CDW phases. To confirm this scenario, we biased the channel with $V_d$ close to 1.8 V ($V_d = 1.81$ V in Figure 4c) and allowed the current to pass the channel for six minutes without applying the gate bias. After reaching the steady-state conditions, we swept the gate voltage and monitored the drain current changes. Interestingly, near $V_g = 38$ V, the current experiences an abrupt and large decrease for about ~ 3 mA, and then goes back to the original level. This change in current is clearly related to the system switching between the IC-CDW and NC-CDW phases. The amount of current drop in Figure 4c is similar to that in Figure 4b. The 1T-TaS$_2$ channel was driven to the IC-CDW phase by the source-drain current at $V_d = 1.81$ V and then switched to the NC-CDW phase by the application of the gate bias of $V_g = 38$ V. The data presented in Figures 4b and 4c proves the possibility of inducing the IC-CDW – NC-CDW phase transition by the gate bias. This is an important capability that was never reported before. In Figure 4d, we show how the source-drain current can be switched by the gate voltage at a small $V_d$ bias, corresponding to the region I of the I-Vs. The changes in the current are due to the effect produced by the gate electric field on C-CDW domains near the gate dielectric. The system remains in the NC-CDW phase during this measurement.

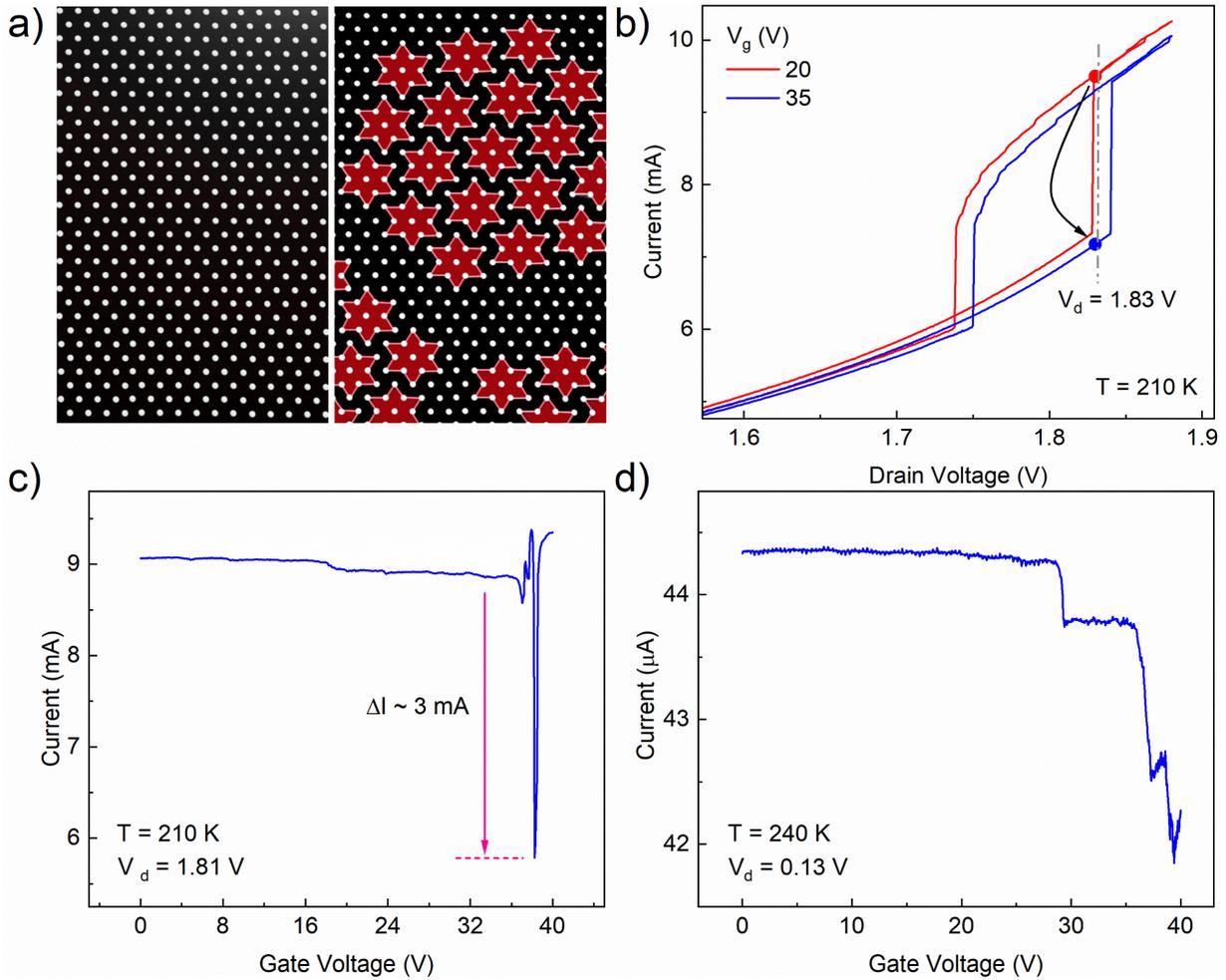

**Figure 4: Switching the charge-density-wave phases with the electrical gate.** (a) Illustration of the IC-CDW and NC-CDW phases. (b) Current as a function of the source-drain voltage at two fixed gate biases, measured at T = 210 K. In this device the transition from IC-CDW to NC-CDW phase, depicted with the red and blue dots, occurs at the voltage close to $V_d$ = 1.83 V. (c) Current as the function of the gate bias at the fixed source-drain voltage $V_d$ = 1.83 V. The current change of $\Delta I \sim 3$ mA, induced by the gate, corresponds to the 1T-TaS₂ channel switching between IC-CDW and NC-CDW phases. (d) Current switching by the gate within the same NC-CDW phase for small, fixed bias of $V_d$ = 0.13 V.

**Discussion**

**CDW de-pinning in quasi-1D and quasi-2D systems.** In the above description of the results, we use the terminology of "de-pinning" of the C-CDW domains to describe the effect of the electric gate on the CDW channel conductance. One should be careful here to understand the difference between the domain de-pinning in quasi-2D materials such as 1T-TaS$_2$ and CDW de-pinning and sliding in "traditional" bulk CDW materials with quasi-1D crystal structures such as NbS$_3$ [2, 49]. In quasi-1D crystals, the entire CDW, *i.e.,* periodic modulation of the charge density, can de-pin and start sliding, resulting in an abrupt increase in current and the appearance of an AC current component [2, 49] (see Figure 5a). In quasi-2D materials, the NC-CDW phase, which separates the IC-CDW and C-CDW phases, consists of the C-CDW islands separated by the metallic IC-CDW phase. The CDW de-pinning in 1T-TaS$_2$ can be understood more like the C-CDW domains becoming softer, loose, start rotating, and changing in size [34, 45, 50-51] (see Figure 5b). The latter contributes to the current fluctuations, observed in the derivative I-Vs or the low-frequency noise spectra [23, 45]. This process does not result in a large increase in the current – the domains start to fluctuate rather than move together [45]. As we established in this work, it also makes the C-CDW domains more susceptible to the gate bias, *i.e.,* electric field. We now use the derivative current characteristics to illustrate the observed gate effects further.

Figure 5c depicts the differential of the drain current with respect to the gate voltage, $dI/dV_g$, as a function of gate bias, $V_g$, at a series of fixed $V_d$ biases. The range of $V_d$ is selected near the on-set of the I-V super-linearity, where the de-pinning of C-CDW domains is expected. The data were obtained at T = 240 K. It can be seen that at $V_d$ = 30 mV and $V_d$ = 130 mV, there are no pronounced changes in $dI/dV_g$. However, increasing the fixed drain biases introduces more fluctuations in the derivative current characteristics. The spikes in the $dI/dV_g$ values shift with increasing $V_d$ and their amplitude increases. We now turn to the differential current characteristic with respect to the source-drain voltage, $dI/dV_d$, as a function of the source-drain voltage $V_d$. Figure 5d, shows one large peak in $dI/dV_d$ around $V_d$ = 1.5 V, which corresponds to $V_H$ and the system transition from NC-CDW to IC-CDW phase in the forward bias sweep. The other peak at $V_d$ = 1.36 V signifies the $V_L$ in the reverse bias sweep. A close look at the small bias region (see inset of Figure 5d), indicates the start of the derivative characteristics at the domain de-pinning threshold voltage $V_T$ =

0.441 V. If we look back at Figure 5c, we notice that the gate derivative characteristics at $V_d = 430$ mV, which is close to $V_T = 0.441$ V, also reveal substantially increased spikes at certain gate biases. At $V_d = 530$ mV, which is above the extracted $V_T$, one observes not only spikes but also increased background fluctuations in $dI/dV_g$. These considerations further confirm that the C-CDW domains in the NC-CDW phase can be affected by both the source-drain bias and the gate voltage.

From the fluctuations, *i.e.,* spikes in the current under the actions of the electric gate we can roughly estimate the physical size of the fluctuator. We assume that the observed changes in the current are associated with some changes in the C-CDW domains that make up the NC-CDW phase [23, 52]. In this case, the change in current, $\Delta I$, is related to the change in the conducting volume, $\Delta V$., *i.e.,* $\Delta I/I = \Delta V/V$. The volume is defined as $V = A \times t \times N$, where $A$ is the channel area, $t$ is the thickness of one atomic plane, and $N$ is the number of the atomic planes in the channel. Taking the smallest and largest normalized change in the current, *($\Delta I/I$) to be* $\sim 8.9 \times 10^{-5}$ - $4.14 \times 10^{-3}$, assuming $N$ to be 25, we obtain the upper and lower bounds of the fluctuator size to be *$(\Delta A)^{1/2} = (AN\Delta I/I)^{1/2}$* $\sim 67$ nm – 455 nm. In the linear region (Figure 3a, region I), the average fluctuator size is $\sim 160$ nm. Naturally, there is ambiguity in the $N$ value since the conduction may not be uniform through the thickness. However, it is clear that we get a reasonable estimate for the fluctuator in the 100 nm order of magnitude. This is consistent with the size of the C-CDW domains in the NC-CDW phase measured by the scanning tunneling microscopy and other methods [10, 37, 53]. These considerations confirm our hypothesis that the gate electric field affects the C-CDW island, and thus induces the spikes in the current, affects the hysteresis and can switch the system between the IC-CDW and NC-CDW phases. The demonstrated electric gating and switching of the CDW phases by electrical field not only contribute to a better understanding of CDW phenomena in quasi-2D material systems but also indicate a path to future technological developments. The demonstrated switching can be potentially utilized in non-volatile memory applications, particularly if the channel area and thickness are downscaled to the 100-nm-scale or below, allowing for control of individual C-CDW domains that exist up to T $\sim 350$ K.

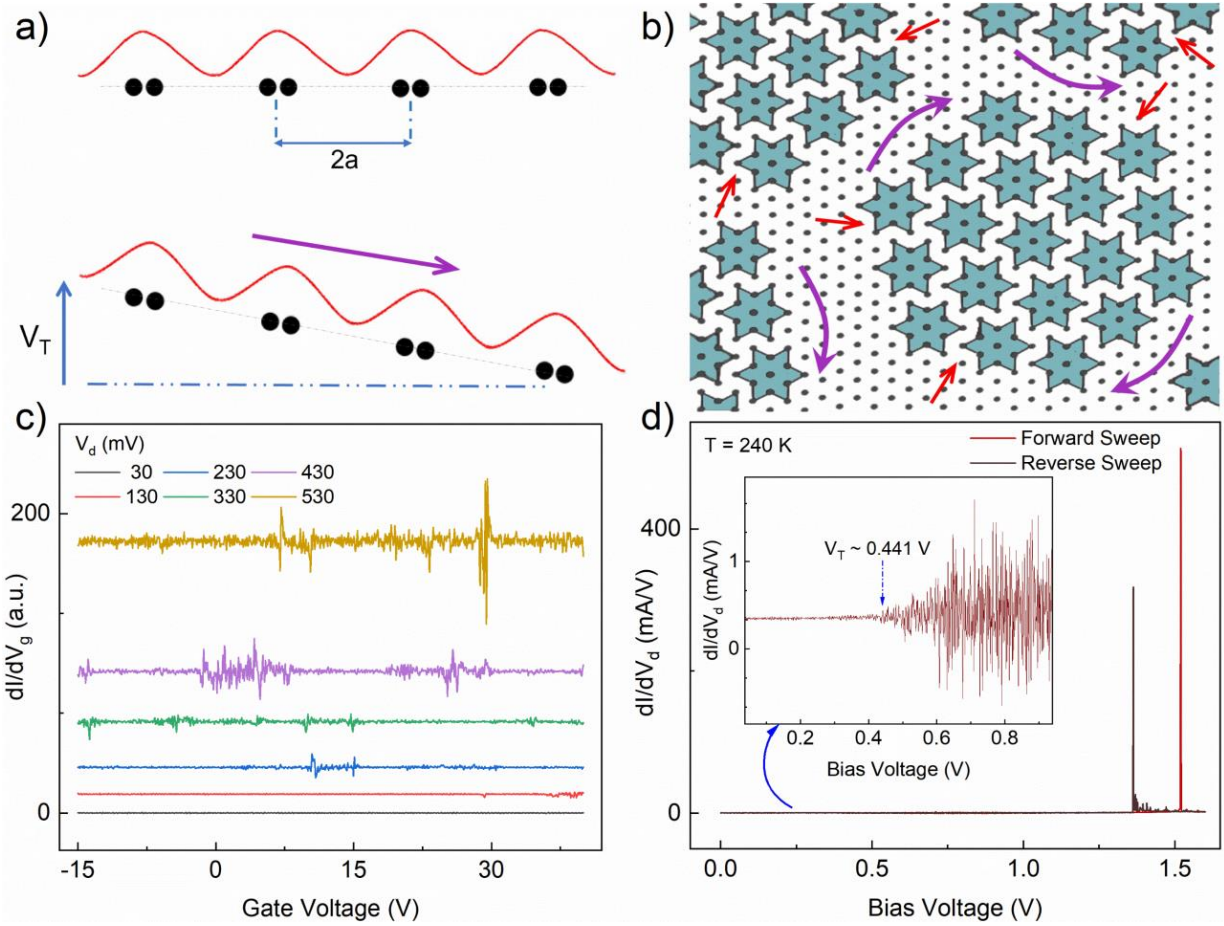

**Figure 5: De-pinning and differential current-voltage characteristics.** Panels (a) and (b) illustrate the difference in the CDW de-pinning in quasi-1D CDW materials and domain "de-pinning" in 2D CDW materials. In quasi-1D chain of atoms, the bias voltage above the threshold voltage, $V_T$, induces the de-pinning and sliding of the whole coherent CDW. In 2D materials the "de-pinning" indicates the onset of the fluctuations, rotation, and movements of the C-CDW domains. (c) Differential current characteristic, $dI/dV_g$, with respect to the gate voltage for fixed $V_d$ values in the linear region I, measured at T = 240 K. (d) Differential current characteristic, $dI/dV_d$, with respect to drain voltages, $V = V_d$. The large spikes at ~1.52 V (red curve) and at ~ 1.36 V (brown curve) are associated with the onset of the hysteresis ($V_H$ and $V_L$) in the forward and reverse sweeps, respectively. The inset depicts the onset of the fluctuations at ~ 0.44 V, within the linear region I. Comparing data in panels (c) and (d), one can see, in both cases, the increase in the fluctuations at drain biases close to the $V_d = 0.44$ V. This fluctuation behavior at the low drain bias indicates the point of de-pinning for the C-CDW domains.

In conclusion, we reported on electrostatic gating of the current and CDW phases in *h*-BN capped 1T-TaS$_2$ heterostructure devices. It is demonstrated that the hysteresis in the source-drain I-V characteristics that accompanies the transition between the NC-CDW and IC-CDW phases can be shifted by applying a gate bias. The gate bias affects the domains of the C-CDW phase as observed from the abrupt changes in the drain current. The transition between the NC-CDW and IC-CDW phases can be induced by both the source-drain current and the electrostatic gate. Since the CDW phases are persistent in 1T-TaS$_2$ at room temperature, one can envision memory applications of such devices when scaled down to the dimensions of individual commensurate domains.

## METHODS

**CDW device fabrication and testing:** High-quality crystals of $1T\text{-}TaS_2$ have been mechanically exfoliated on top of polydimethylsiloxane (PDMS) layers in a dry environment (with relative humidity of < 19%). Thin films of $1T\text{-}TaS_2$ were then transferred onto pre-selected $h$-BN layers on $Si/SiO_2$ substrate through the all-dry transfer method. This step is immediately followed by capping the $1T\text{-}TaS_2$ with another few-layer aligned $h$-BN film. The devices being capped/encapsulated with $h$-BN will preserve the CDW state longer and give better protection from oxidation and environmental degradation. The patterning of electrodes was performed through electron-beam lithography. The $h$-BN layers covering the electrode contacts were removed with high precision using the atomic layer etching technique. For contact metal deposition, we used electron-beam evaporation and deposited 10/100 nm of Ti/Au metals. The fabricated devices had dimensions of 1 to 3 µm channel length and 1 to 3.5 µm channel width. The gate measurements at various temperatures were conducted in the Lakeshore cryogenic probe station TTPX and further analyzed with a semiconductor analyzer Agilent B1500.

## Acknowledgments

The work at UC Riverside was supported, in part, by the U.S. Department of Energy Office of Basic Energy Sciences under contract No. DE-SC0021020 "Physical Mechanisms and Electric-Bias Control of Phase Transitions in Quasi-2D Charge-Density-Wave Quantum Materials". A.A.B. was supported by the Vannevar Bush Faculty Fellowship from the Office of Secretary of Defense (OSD), under the Office of Naval Research (ONR) contract N00014-21-1-2947. A.R. and S.R. work was partially supported by the "International Research Agendas" program of the Foundation for Polish Science co-financed by the European Union under the European Regional Development Fund No. MAB/2018/9.

## Author Contributions

A.A.B. and S.R. conceived the idea. A.A.B coordinated the project and led the experimental data analysis and manuscript preparation. A.R., S.R., and F.K. contributed to the data analysis. N.S. synthesized and characterized bulk crystals of 1T-TaS$_2$, supervised by T.T.S. M.T. fabricated devices, conducted measurements, and contributed to the data analysis. J.B. assisted with device fabrication and contributed to data analysis. A.A.B. and M.T. wrote the manuscript. All authors contributed to the manuscript editing.

## Supplemental Information

The supplemental information is available on the journal website for free of charge.

## The Data Availability Statement

The data that support the findings of this study are available from the corresponding author upon reasonable request.